# Exploring Spectrum Sensing Techniques in Cognitive Radio Systems Using Time-Domain Symbol Cross-correlation

Ahmed Temtam, Dimitrie Popescu.
atemt001@odu.edu, DPopescu@odu.edu
Department of Electrical and Computer Engineering, Old Dominion University
Norfolk, Virginia, USA

**Abstract**-In order to enable spectrum sharing, spectrum sensing plays a crucial role in wireless communication. The challenges in wireless spectrum require collaboration among stakeholders to devise innovative solutions. This research explores the use of a Cognitive Radio (CR) system that employs a Time-Domain Symbol Crosscorrelation (TDSC) based spectrum sensing algorithm. WiMAX and LTE standards are utilized as case studies to demonstrate the efficacy of the TDSC method. The study presents theoretical and simulation results and also suggests future research to investigate the performance of the TDSC method in WiMAX and LTE systems. Additionally, this study compares the spectrum sensing capabilities of WiMAX and LTE.

I. INTRODUCTION

Wireless technology has become ubiquitous in modern telecommunication systems, catering to a wide range of data communications needs. With the growing popularity of mobile devices and wireless internet, there is a pressing demand for faster wireless services to keep up with evolving software applications. The WiMAX standard, based on IEEE 802.16, initially focused on the 10 to 66 GHz band. Later, the 802.16a amendment extended the standard to cover the 2 to 11 GHz band [1]. The 802.16-2004 version was further enhanced by the 802.16e-2005 amendment, which introduced scalable Orthogonal Frequency-Division Multiple Access (OFDMA) as the access method. OFDMA, based on Orthogonal Frequency Division Multiplexing (OFDM), is a digital data encoding technique that utilizes multiple carrier frequencies. WiMAX does not have a globally standardized licensed spectrum; instead, the WiMAX Forum has specified three licensed spectrum bands (2.3 GHz, 2.5 GHz, and 3.5 GHz) to promote interoperability and reduce costs [2].

Channel sensing in WiMAX can be achieved by either injecting pilot tones into specific time slots of OFDM symbols or injecting pilot tones into all OFDM symbols. If a cognitive radio (CR) operator detects the presence of a primary user in the allocated spectrum, it must switch to an available alternative band. Similarly, if a secondary user detects an unlicensed operator, it can switch to another available band or negotiate spectrum sharing with the existing operator. Therefore, CR signals require a reliable source identification feature to facilitate active and dependable spectrum sharing.

In contrast, Long Term Evolution (LTE), a mobile telecommunications technology standardized by 3GPP, represents a significant advancement from 3G UMTS and CDMA2000 towards 4G networks. LTE meets the increasing requirements for data rates, capacity, and latency [3][4]. As the primary standard for 4G communication systems, LTE supports broadband applications with data rates of up to 100 Mbps in the downlink and 50 Mbps in the uplink, utilizing a bandwidth of up to 20 MHz [3][4].

At the physical layer, LTE employs OFDM, which has emerged as a favored radio access scheme due to its simplicity of implementation and scalability [1]. OFDM is also utilized in other wireless standards such as IEEE 802.11 for wireless local area networks (WLAN) and IEEE 802.15 for short-range and personal area networks (WPAN), as well as in current digital television standards [5][6]. Furthermore, OFDM has been proposed for use in future wireless systems employing cognitive radio technology, allowing unlicensed devices to access licensed spectrum under strict constraints [7].

With the anticipation of significant growth in mobile data traffic and the need for more efficient spectrum utilization, recent studies have explored the use of LTE systems in heterogeneous networks that incorporate licensed LTE frequencies alongside unlicensed frequencies like WLAN bands or TV white spaces [8][9].

The motivation of this paper is to address the aforementioned challenges by proposing an adaptive OFDM system that incorporates a spectrum sensing algorithm based on Time Domain Symbol Cross-correlation (TDSC) of two OFDM symbols. Incorporating TDSC-based spectrum sensing into adaptive OFDM systems can significantly bolster the efficiency, accuracy, and adaptability of wireless communication networks. As the demand for spectral resources continues to surge, such innovative solutions will become increasingly essential in ensuring the seamless and efficient operation of next-generation communication systems.

## II. ORTHOGONAL FREQUENCY DIVISION MULTIPLEXING (OFDM)-BASED COGNITIVE RADIO.

Cognitive radio offers a promising solution to address the issue of spectral congestion by introducing the concept of opportunistic spectrum utilization. By operating as secondary systems alongside primary (or licensed) systems, cognitive radios have the ability to detect and access unused spectrum. In the realm of wireless communication systems, Orthogonal Frequency-Division Multiplexing (OFDM) has gained significant popularity and is widely utilized due to its numerous advantages [10]. OFDM has proven to be successful in various wireless standards and technologies. Its suitability for cognitive radio systems is evident as it provides a reliable, scalable, and adaptable technology for air interface. Extensive research indicates that OFDM is a viable candidate for implementing cognitive radio concepts.

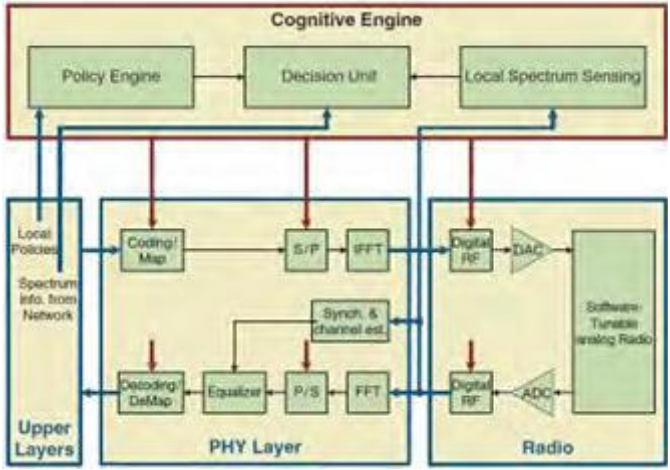

Figure 1. Block diagram of a generic OFDM transceiver [7].

In order for cognitive radio (CR) to fulfill its objectives, the Physical Layer (PHY) needs to possess high adjustability and flexibility. Orthogonal Frequency-Division Multiplexing(OFDM) stands out as one of the most commonly used technologies in modern wireless communication systems [11], offering the potential to meet the aforementioned requirements of CR with minimal modifications. OFDM employs multicarrier transmission, where the spectrum is divided into sub-bands that are modulated with orthogonal subcarriers. This approach eliminates the need for equalizers, thereby simplifying the receiver design. The intelligent structure of OFDM has proven effective in various wireless technologies. It is expected that OFDM will similarly excel in enabling cognitive radio concepts, as it provides a well-established, scalable, and adaptive technology for air interface. Figure 1 illustrates a basic block diagram of a simple OFDM system.

## III. SYSTEM MODEL

### A. WiMAX System Model

The primary goal of the WiMAX network model is to establish an IP-based infrastructure that offers scalable data capacity, open access to innovative applications and services, and improved quality of service (QoS) and mobility. The IEEE 802.16 standards define the specifications for the Physical and Link Layer configurations that govern the interactions between mobile stations (MSs) and base stations (BSs).

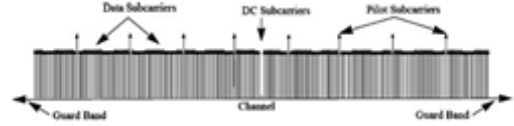

Figure 2. OFDM frequency description [12].

### B. LTE System Model

LTE systems currently deployed globally predominantly utilize Frequency Division Duplex (FDD) for implementing the downlink (DL) and uplink (UL) channels between a provider base station and a mobile subscriber. Consequently, two distinct frequency bands are employed for communication between the mobile subscriber terminal and the base station.

To comprehend the periodic characteristics of the pilot information found in LTE signals, we can begin by examining the frame structure defined by the LTE standard. This structure is divided into 20 individual slots, each lasting for 0.5 ms, as depicted in Figure 3 for the FDD DL channel. Each slot comprises NDL symb OFDM symbols. The exact number of symbols depends on the length of the cyclic prefix (CP) and the parameters defining the duration of useful symbols in the OFDM signal [3], [4].

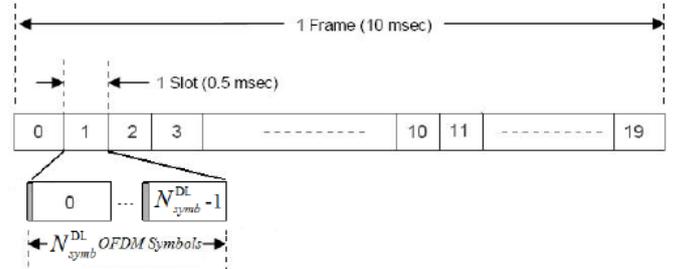

Figure 3. FDD DL frame structure in the LTE systems. [12].

*In the LTE frame* structure, each slot is visualized as a two-dimensional grid, as illustrated in Figure 4. It consists of $N_{\text{symb}}^{\text{DL}}$ OFDM symbols in the time domain and $N = N_{\text{RB}}^{\text{DL}} N_{\text{sc}}^{\text{RB}}$ subcarriers in the frequency domain[1]. The LTE system employs resource blocks (RBs) for defining these slots, with $N_{\text{RB}}^{\text{DL}}$ indicating the number of RBs and $N_{\text{sc}}^{\text{RB}}$ representing the number of subcarriers within a RB. Specifically, a resource block comprises a consecutive set of $N_{\text{symb}}^{\text{DL}}$ OFDM symbols in the time domain and $N_{sc}^{RB}$ consecutive subcarriers in the frequency domain. For LTE signals, $N_{sc}^{RB}$ is equal to 12 for subcarrier spacing $\Delta f$ = 15 kHz and 24 for $\Delta f$ = 7.5 kHz.

---

[1] Here, *N* denotes the total number of subcarriers in an OFDM symbol.

The resource grid, depicted in Figure 4, is further divided into resource elements, which are the smallest units of the grid.

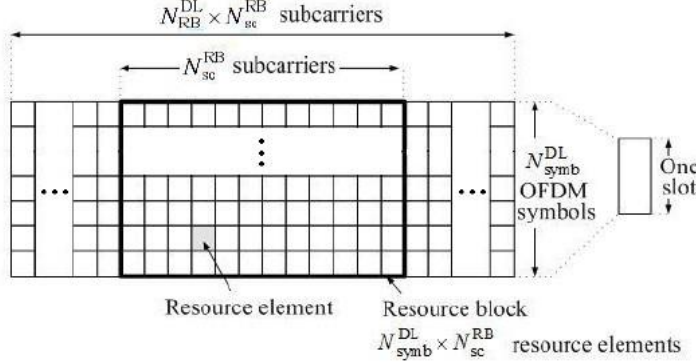

Figure 4. Slot structure and resource grid in the FDD DL frame [12].

A resource block consists of NsymbDL ×NscRB resource elements. In the LTE standard, the pilot information is embedded in the resource blocks of the transmission frame and is referred to as reference signals (RS). These reference signals serve the purpose of channel estimation, synchronization, and cell search/acquisition, as depicted in Figure 5 [3], [4].

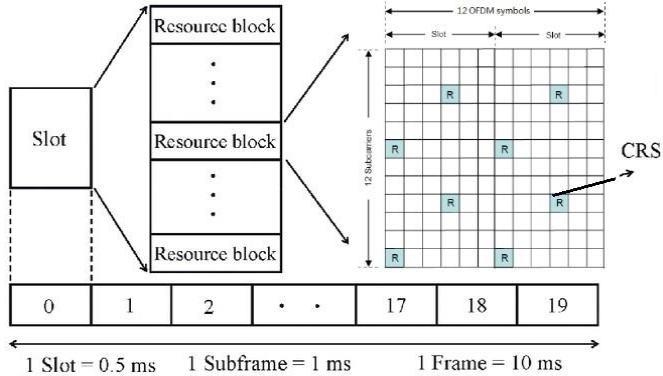

Figure 5. Resource element mapping of pilot information in LTE signals.

In the LTE network, each cell is assigned a reference signal (RS) that serves as a cell identifier. Consequently, the RS is repeated in each downlink frame. The RSs are distributed across the resource elements and are typically transmitted on specific subcarriers of one or two non-consecutive symbols in each slot. Figure 5 illustrates the distribution of the cell specific RS for long cyclic prefix (CP) over one resource block and two consecutive slots, where each slot contains N symb DL = 7

OFDM symbols and the resource block has NscRB = 12 subcarriers. In this example, the cell-specific RS is transmitted on the first and seventh subcarriers of the first OFDM symbol, and on the fourth and tenth subcarriers of the fourth OFDM symbol in each slot.

It is worth noting that the structure of the uplink (UL) slots is similar to the downlink (DL) slots, although there may be differences in the reference symbol configuration, robustness, and physical multiplexing of the UL channel. Due to space limitations, a detailed description of the UL slot structure is not provided here.

In the LTE network, each cell is assigned a reference signal (RS) that serves as a cell identifier. Consequently, the RS is repeated in each downlink frame. The RSs are distributed across the resource elements and are typically transmitted on specific subcarriers of one or two non-consecutive symbols in each slot. Figure 5 illustrates the distribution of the cell specific RS for long cyclic prefix (CP) over one resource block and two consecutive slots, where each slot contains NsymbDL = 7

OFDM symbols and the resource block has NscRB = 12 subcarriers. In this example, the cell-specific RS is transmitted on the first and seventh subcarriers of the first OFDM symbol, and on the fourth and tenth subcarriers of the fourth OFDM symbol in each slot.

It is worth noting that the structure of the uplink (UL) slots is similar to the downlink (DL) slots, although there may be differences in the reference symbol configuration, robustness, and physical multiplexing of the UL channel. Due to space limitations, a detailed description of the UL slot structure is not provided here.

IV. SENSING OFDM SIGNALS

One of the most challenging tasks in cognitive radio (CR) is the development of a scheme to sense the availability of frequency bands and adjust the parameters of the communication system accordingly, even when the received signal at the sensing element of the CR has a low signal-to-noise ratio (SNR). The sensing system must be able to operate reliably in low SNR conditions [13][14][15] [16].

The TDSC algorithm leverages the cross-correlation between the OFDM symbols to estimate the presence or absence of primary users in the spectrum. This approach allows for reliable spectrum sensing even in challenging SNR conditions.

In this paper, the proposed sensing algorithm is based on the Time-Domain Symbol Cross-correlation (TDSC) of two OFDM symbols. The algorithm is designed to work effectively even when the length of the time-invariant channel, denoted as $L$, is longer than the length of the OFDM symbols. By considering the $n$th sample of the $l$th OFDM symbol [17], the algorithm can be expressed as follows:

$$x_l[n] = e^{j(2\pi f_\Delta n/N + \theta_l)} \cdot \frac{1}{N}\sum_{k=0}^{N-1} H[k]X_l[k]e^{j2\pi kn/N} + w_l[n]$$
(1)

where
$L$ is the length of the Cyclic Prefix (CP).
$\Delta f$ is the carrier frequency offset normalized to the subcarrier spacing. $\theta$ is the initial phase of the $l$th OFDM symbol.

$M = N + L$ is the length of an OFDM symbol.
$N$ is the number of subcarriers.
$X_l[k]$ is the data symbols at the $k$ th subcarrier of the $l$th OFDM symbol.
$H[k]$ is the complex channel gain of the $k$th subcarrier. $w_l[n]$ is a sample of a complex additive white Gaussian noise (AWGN) process.
$w_l[n]$ assumed to be a circularly symmetric complex Gaussian random variable, variance of $\sigma_w^2/N$.
Under the assumption that $P_a$, the sets of all likely pilot tone locations for the communicated OFDM symbols. The TimeDomain Symbol Cross-correlation (TDSC) defined as

$$R(l,m) = \frac{1}{N}\sum_{n=0}^{N-1} x_l[n]x_m^*[n] \quad (2)$$

Where $l$th and $m$th OFDM symbols have the same pilot tone positions. After a straightforward calculations, that shown in[17].

$$R(l,m) \cong e(l-m) \cdot \frac{\rho^2}{N^2}\sum_{k\in \mathbf{P}_a} |H[k]|^2$$
$$+ \frac{1}{N}\sum_{k=0}^{N-1} w_l[n]w_m^*[n] . \quad (3)$$

The Eq.3. contains of a static term and a noise term.

## V. SENSING ALGORITHM

According to [17], the symbol index difference between two OFDM symbols is denoted as v = l – m. When the symbol index difference is equal to v, it means that the two symbols have the same pilot tone configuration. Additionally, the accumulated Time-Domain Symbol Cross-correlation (TDSC) function is denoted as C(v). This function represents the accumulated cross-correlation between the OFDM symbols with a symbol index difference of v.

$$C(v) = \frac{1}{S_v}\sum_{v=m-l} R(l,m)$$
$$= e(v)\frac{\rho^2}{N^2} \cdot \frac{1}{A}\sum_{a=0}^{A-1}\sum_{k\in \mathbf{P}_a} |H[k]|^2$$
$$+ \frac{1}{N}\sum_{v=m-l}\sum_{k=0}^{N-1} w_l[n]w_m^*[n] . \quad (4)$$

Where
$S_v$ is the number of $R(l,m)$ that are accrued and supplemented.
$S_v$ is designated to be an integer compound of $A$.
It's clear from (4) the mean of $C(v)$ is unaffected. Nevertheless, the variance of the second term which is noise term in $C(v)$ is contrariwise related to $S_v$. Thus, even though, the accumulated number of $R(l,m)$, $S_v$ is increased, the noise term in $C(v)$ will decreased. As a result, it could be achieved spectrum sensing in signil cantly low SNR situations. According to [17] $C(v)$

$$C(v) = e(v)\Lambda + \zeta(v). \quad (5)$$

where

$$\Lambda = \frac{\rho^2}{N^2} \cdot \frac{1}{A}\sum_{a=0}^{A-1}\sum_{k\in \mathbf{P}_a} |H[k]|^2 \quad (6)$$

$\Lambda$ is the average received signal power in the pilot tone positions divided by $N^2$.
$\zeta(v)$ is a circularly symmetric complex Gaussian random variable.
The possible decisions with binary hypotheses testing are:

$$H_0 : C(v) = \zeta(v).$$
$$H_1 : C(v) = e(v)\Lambda + \zeta(v). \quad (7)$$

where $H_0$ noise only. $H_1$ is signal and noise existing. The likelihood ratio function is expressed as

$$L(\mathbf{C}) = \frac{p(\mathbf{C};H_1)}{p(\mathbf{C};H_0)}. \quad (8)$$

After performing certain direct calculations, the decision statistic of the Neyman-Pearson (NP) test [18] can be expressed as follows:

$$T_{NP} = |C(v)|. \quad (9)$$

## VI. SPECTRUM SENSING FOR WIMAX OFDM SYSTEMS

### A. Probability of Misdetection

The probability of misdetection, denoted as $P_{MD}$, represents the probability that the secondary user incorrectly decides that a primary user is not present in a given channel, even though the primary wireless system is actually utilizing that channel. The probability distributions for hypotheses $H_0$ and $H_1$ in TDSC spectrum sensing algorithms are assumed to be circularly symmetric complex Gaussian. The variances of the distributions for $H_0$ and $H_1$ are denoted as $\sigma_{H\,0}^2$ and $\sigma_{H\,1}^2$, respectively, while $\mu^2_{H_0}$ represents the mean of the distribution for $H_1$. In the context of hypothesis $H_0$, a Rayleigh distribution is used for $|T_{NP}|$.

On the other hand, the probability of false alarm, denoted as $P_{FA}$, indicates the probability that the secondary user incorrectly decides that a primary user is present in a certain channel, even though the primary wireless system is not utilizing that channel. According to [17], the matching threshold $\gamma$ can be determined as follows:

$$\gamma = \sqrt{-\sigma_{H_0}^2 \ln P_{FA}}. \quad (10)$$

For sake of simplicity it's been used a single-path channel, the probability of misdetection $P_{MD}$ is expressed as

$$P_{MD} = 1 - Q_{x_2'^2(\lambda)}\left(\frac{\gamma^2}{\sigma_{H_1}^2}\right). \quad (11)$$

The function $Q_{x_2'^2(\lambda)}(x)$ represents the right-tail probability of the non-central chi-squared distribution with two degrees of freedom, denoted as $x_2'^2(\lambda)$.

## VII. FEATURES OF LTE/WIMAX

Both WiMAX and LTE systems utilize Orthogonal Frequency Division Multiple Access (OFDMA) in the downlink (DL) with higher-order modulation and coding, resulting in similar peak performance for the same modulation and code rate. They both support Frequency Division Duplex (FDD) and Time Division Duplex (TDD) with channel bandwidths up to 20 MHz. Additionally, both systems offer support for higher order Multiple-Input Multiple-Output (MIMO) antenna solutions and aim to reduce latency [19].

Furthermore, both WiMAX and LTE are all-IP, packet-based technologies with a packet network core. This makes them well-suited for handling burst data traffic and providing good support for Voice over IP (VoIP). They both employ OFDMA, a form of Frequency Division Multiplexing (FDM) where the subcarriers are made orthogonal to each other [20]. This orthogonal arrangement allows for packing more subcarriers into the available spectrum, leading to higher spectral efficiency. However, this also results in larger symbol sizes, which helps in mitigating Inter-Symbol Interference (ISI) and reduces the need for complex adaptive equalization techniques used in wideband systems with a single carrier [21]. OFDMA is robust against frequency selective burst errors and narrowband interference.

In both LTE and WiMAX, the connection is organized in both time and frequency domains, with multiple connections sharing multiple carriers. This sharing can be periodically adjusted to maximize system performance [22]. Some additional features associated with LTE and WiMAX include:

1. Sub-channelization and permutation: Within the assigned spectrum, some subcarriers are used for data, while others are utilized as guard bands and pilots. Data carriers and pilots are periodically selected for different sub-channels, resulting in frequency hopping. This reduces interference and improves system capacity [21].

2. Partial use of subcarriers (PUSC): Subcarriers are subdivided into clusters, and only specific clusters can be utilized in each cell. This approach reduces interference from neighboring cells and improves overall performance.

3. Fractional frequency reuse (FFR): FFR is employed to manage interference. Terminals located near the cell center use all frequencies, while those near the cell boundary use different frequencies to minimize inter-cell interference [23].

Both LTE and WiMAX utilize a variant of OFDMA called scalable OFDMA (SOFDMA), which dynamically adjusts the number of subcarriers based on the allocated bandwidth. This ensures that the Doppler effect on performance remains consistent for mobile users [24].

Both LTE and WiMAX employ Adaptive Modulation and Coding (AMC) for link adaptation. This allows the system to adapt the modulation scheme and coding rate based on the current signal conditions, ensuring a suitable Quality of Service (QoS) and extending the range for users experiencing lower Signal-to-Noise Ratio (SNR) [25]. Users with enhanced SNR can be assigned higher modulation schemes to achieve higher data rates and increase capacity. The combination of AMC with multicarrier OFDM yields additional benefits, as adjusting a narrowband channel to noise settings is more efficient than adapting to averaged noise in a wideband channel [26].

Both LTE and WiMAX also incorporate Hybrid Automatic Repeat Request (HARQ) for error detection and correction, as well as multiple antenna technologies to further enhance performance and data rates. In the case of WiMAX, the 4G version introduces various enhancements at the physical layer, including

## VIII. TECHNICAL DIFFERENCES

There are several technical similarities between LTE and WiMAX in terms of architecture and objectives. Both LTE and WiMAX utilize Orthogonal Frequency Division Multiple Access (OFDMA) with a flat IP architecture, aiming to meet or surpass the requirements of IMT-Advanced. They also employ supporting technologies in parallel [27]. However, there are some technical differences between WiMAX and LTE, including the following:

1. Duplex Mode: Both LTE and WiMAX support Time Division Duplex (TDD) and Frequency Division Duplex (FDD). However, FDD has been the primary focus of telecom corporations and has remained consistent across different generations. LTE has gained recognition as the evolution path of synchronous CDMA, while WiMAX has focused on TDD. In the future, WiMAX implementations may likely move towards TD-LTE [28].

2. Spectrum: LTE operates in licensed IMT-2000 bands, such as 700, 900, 1800, 2100, and 2600 MHz, while WiMAX operates in both licensed and unlicensed bands, such as 2.3, 2.5, 3.5, and 5.8 GHz. This gives LTE an advantage in terms of availability in desired low-frequency bands, making it suitable for public wide area networks. Some operators have started exploring LTE in the WiMAX bands they already possess [25].

3. Intercarrier Spacing: LTE uses a standard intercarrier spacing of 15 kHz, while WiMAX uses 10.94 kHz. A larger intercarrier spacing provides better protection against Doppler spread. LTE can support mobility speeds up to 350 km/h, whereas WiMAX can handle speeds of around 120 km/h [29].

4. Access Technology: LTE employs Orthogonal Frequency Division Multiple Access (OFDMA) for its downlink, while it uses Single Carrier Frequency Division Multiple Access (SCFDMA) for the uplink. SC-FDMA reduces Peak-to-Average Power Ratio (PAPR) by 3-5 dB, resulting in improved uplink coverage or throughput at cell borders. WiMAX uses OFDMA for both uplink and downlink. OFDMA, as well as SC-FDMA, are suitable for broadband systems due to their robustness against multipath signal propagation [21], [30].

In addition to these technical differences, other factors such as regional regulations, operator preferences, and control factors also influence the choice between LTE and WiMAX. WiMAX was introduced earlier and has a TDD nature, which offers flexibility in sharing the time frame between uplink and downlink. This made it initially more suitable for data as a wireless alternative to wired DSL. However, the emergence of TD-LTE, a TDD version of LTE with a single band for operation, eliminates one of WiMAX's core advantages over LTE [31], [32].

It should be noted that LTE's design appears to be more focused on mobility, data throughput, and capacity, but these features alone may not be the sole determinants for choosing one technology over the other [24]. Other factors, including regional, operator, and regulatory considerations, also play a significant role in technology selection.

| Characteristic | 3GGP Track | IEEE 802.16 Track |
|---|---|---|
| All IP vs Circuit Switched | Started Circuit switched, moved to half IP (2.5/3 G) and finally All IP (LTE) | All IP from the beginning |
| Architecture | Centric architecture, gradually moving to flat architecture | Flat architecture from the beginning |
| Mobility | Started voice centric gradually moved to data centric | Started as data centric gradually serving voice |
| Mode of operation | FDD is the main mode with increased interest in TDD recently | TDD mode mainly |
| Access Technology | Different access technologies like TDM/FDM and Spread Spectrum before heading to OFDMA in LTE | OFDMA was considered at early stages by IEEE 802.16 standards |
| Spectrum | Lower licensed bands | Higher licensed and unlicensed bands |
| Target | Targeted wide coverage and ubiquitous service | Targeted spotty dedicated coverage. Failed to provide ubiquitous coverage later on |

Table I
COMPARISON OF 3GPP(LTE) AND IEEE 802.16 (WIMAX)[33].

The WiMAX standards, based on the IEEE standards, are modular and separate, offering high performance. However, the 4G version of WiMAX lacks support for legacy 3GPP devices, meaning there are no handover capabilities. On the other hand, the 3GPP standards have provided a clear evolution path towards LTE. This has resulted in operators worldwide who have already deployed their networks based on 3GPP standards finding it commercially advantageous to transition to LTE. The migration to LTE offers easy upgrades and allows operators to reuse their existing spectrum from outdated technologies like 2G for more efficient LTE deployments [34]. Table 3 presents a comparison between the 3GPP track leading to LTE and the IEEE 802.16 track leading to the current version of WiMAX.

## IX. SIMULATION AND DISCUSSION

In the following section, we present simulation results using TDSC spectrum sensing for WiMAX and LTE OFDM signals.

### A. Simulation

For the simulations, the signal-to-noise ratio (SNR) was varied from -23 dB to -14 dB, with a fixed probability of false alarm $P_{FA}$ set to 0.01. The threshold for detection was determined based on Equation (10).

We conducted simulations using the pilot pattern configurations specified in [35] [36] for WiMAX OFDM signals. The parameters chosen for the simulation are as follows: the FFT size and the number of subcarriers, denoted as M, are set to 1024; the simulated signals have a bandwidth of 5 MHz on each side; the cyclic prefix duration $T_{cp}$ for mobile WiMAX is set to 1/8.

The data subcarriers are modulated using 16-QAM with unit variance of the signal constellation, following the IEEE 802.16e standard [37]. The pilot subcarriers in mobile WiMAX are also modulated accordingly. In the simulations, the number of symbols in the uplink subframes is set to 35, while in the downlink subframes it is set to 12. The durations of the RTG and TTG are specified as 60 $\mu s$ and 107.225 $\mu s$, respectively, according to the WiMAX standard [35]. To ensure that the post-fix does not exceed the predefined cyclic prefix, the transmitter window of the OFDM employs a maximum roll-off factor of 0.1. The sampling frequency is set to 8.4 MHz, and the signal is subjected to a phase offset $\varphi$ uniformly distributed in the range of $[-\pi, \pi]$, as well as a carrier frequency offset of 0.5.

These simulation parameters allow us to evaluate the performance of TDSC spectrum sensing for WiMAX and LTE OFDM signals.

In the simulations, we considered various channel environments, including AWGN (Additive White Gaussian Noise), multipath Rayleigh fading, and multipath Ricean fading channels. Specifically, we utilized the ITU-R normal and vehicular A fading channels. The maximum delay spread for the ITUR normal channel is 410 ns, while for the vehicular A fading channel, it is 2.51 ÎŒs [38]. The maximum Doppler frequencies for the ITU-R pedestrian and vehicular A fading channels are set at 7.28 Hz and 145.69 Hz, respectively.

At the receiver, a filter is employed to remove out-of-band noise, and the signal-to-noise ratio (SNR) is modulated at the output of this filter. The performance evaluation focuses on the probability of missed detection ($P_{MD}$) and the thresholds used in the TDSC spectrum sensing tests, with a fixed probability of false alarm ($P_{FA}$) set to 0.01 and a sensing time of 50 ms. The cyclic prefix (CP) ratios are set to values of 1/4 and 1/8.

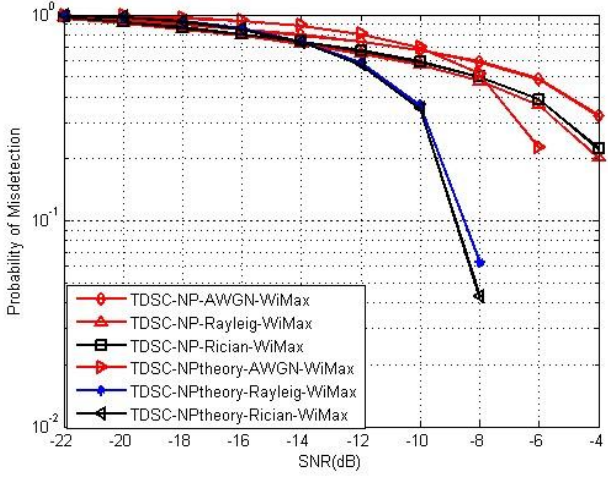

Figure 6. Performance of the TDSC-CP-WiMax method and the reference value for $P_{FA}$=0.01, CP length = 1/4

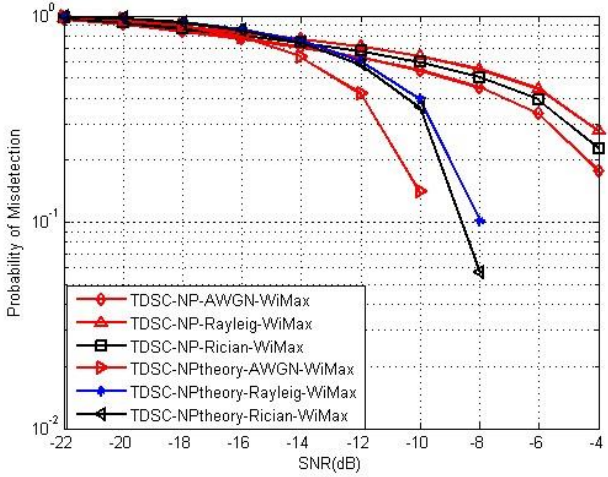

Figure 7. Performance of the TDSC-CP-WiMax method and the reference value for $P_{FA}$=0.01, CP length = 1/8

Figure (6,7) shows the sensing performance under the multipath channel conditions, specifically AWGN and Ricean fading, where the performance remains relatively stable. However, under the multipath Rayleigh fading channel, there is a slight change in the sensing performance. Figure (8) displays a set of receiver operating characteristic (ROC) curves for the TDSC-NP method at different SNR levels. Considering both the TDSC-NP method and the channel environments, we selected a CP ratio of 1/4 and the multipath AWGN scenario.

These simulation environments enable us to assess the performance of TDSC spectrum sensing in different channel conditions and evaluate its robustness and reliability.

*B. Experimental Results and Discussion*

This section presents the simulation results and underlying assumptions used to illustrate the accuracy of the method for evaluating spectrum sensing performance. Extensive simulations were conducted with the following setup:

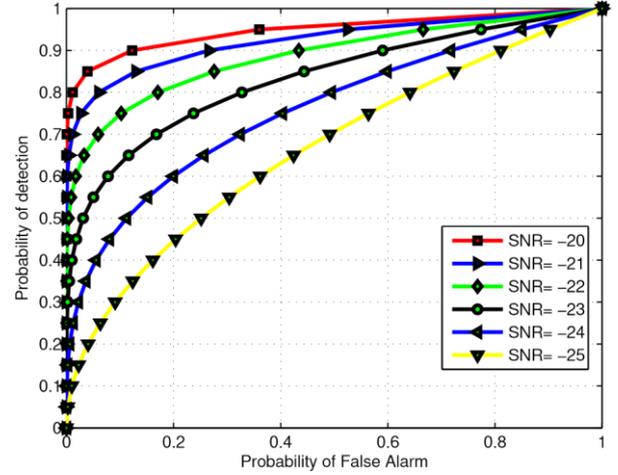

Figure 8. Family of ROC curves for TDSC-CP-WiMax at different levels of SNR, CP length = 1/4 and sensing time = 50 ms.

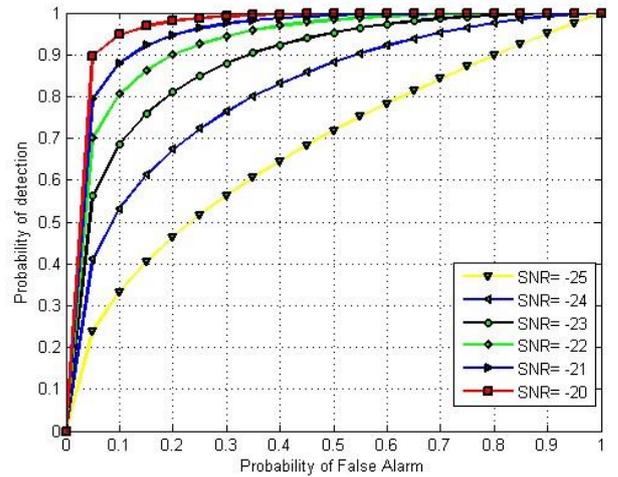

Figure 9. Family of ROC curves for TDSC-CP-LTE at different levels of SNR, CP length = 1/4 and sensing time = 50 ms.

- The signal-to-noise ratio (SNR) was varied from -23 dB to -14 dB. - The probability of false alarm ($P_{FA}$) was set to 0.01, and the threshold was determined using Equation (10). - TDSC spectrum sensing was applied to the pilot pattern assemblies of WiMAX OFDM signals, using parameters defined in [39]. - For the simulations, the OFDM WiMAX physical layer parameters were chosen as follows: FFT size (M) equal to 1024, signals simulated with a 5 MHz double-sided bandwidth, and a mobile WiMAX signal cyclic prefix duration ($T_{cp}$) set to 1/4 and 1/8. - The transmission utilized QAM modulation with 16 points, and the data subcarriers were modulated using a signal constellation with unit variance. - The pilot subcarriers in mobile WiMAX were modulated according to the IEEE 802.16e standard [40]. - The number of symbols in the uplink subframes was 35, while for downlink subframes, it was 12. The RTG duration was 60s, and the TTG duration was 107.225s [38]. - The rolloff factor of the OFDM transmitter window was set to a maximum value of 0.1 to ensure the postfix did not exceed the predefined cyclic prefix. - The sampling frequency was modulated to 8.4 MHz, and the signal was subjected to a phase offset uniformly distributed in the range of [-π, π]. The carrier frequency offset was set to 0.5.

The simulation environments included the following channel models: AWGN, multipath Rayleigh fading, multipath Ricean fading, ITU-R normal fading, and vehicular A fading. The maximum delay spread for the ITU-R normal channel was 410 ns, while for the vehicular A fading channel, it was 2.51s [38]. The maximum Doppler frequencies for the ITU-R pedestrian and vehicular A fading channels were set at 7.28 Hz and 145.69 Hz, respectively.

At the receiver, a filter was utilized to eliminate out-ofband noise, and the SNR was modulated at the output of this filter. The performance evaluation focused on the probability of missed detection ($P_{MD}$) and the thresholds used in the TDSC spectrum sensing tests, with a fixed $P_{FA}$ of 0.01 and a sensing time of 50 ms. The cyclic prefix (CP) ratios were set as [1/4, 1/8]. Figure (9,10) shows the sensing performance under the multipath channel (AWGN and Ricean) condition does not reduce much whereas the sensing performance under the multipath channel Rayleigh the sensing presentations of TDSC-NP method and channel environments, we choose a CP ratio of 1/4 and multipath AWGN.

Figure (9) illustrates the performance comparison of WiMAX and LTE using subcarrier TDSC-NP schemes across different channel conditions. The simulation results indicate that both WiMAX and LTE exhibit better performance in Rayleigh and Rician fading channels compared to AWGN. Additionally, WiMAX demonstrates slightly better performance than LTE across all channel conditions.

In Figure (10), the probability of missed detection ($P_{MD}$) performance of the TDSC system is shown for AWGN, Rayleigh, and Rician channels using a cyclic prefix (CP) length of 1/8 for both WiMAX and LTE. For the AWGN channel, both simulation and theoretical results exhibit similar performance for both techniques. However, in the case of the Rayleigh channel, there is a noticeable difference in the $P_{MD}$ performance between the simulation and theory. Moreover, it is observed that the behavior of the AWGN, Rayleigh, and Rician channels is quite similar for both WiMAX and LTE.

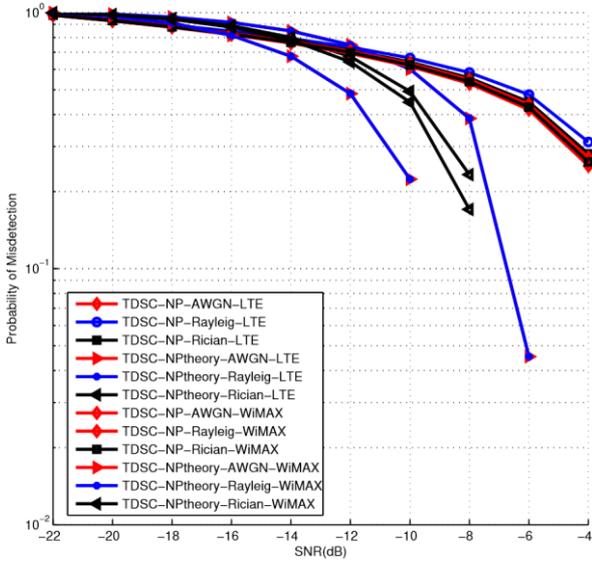

Figure 10. Performance Comparison of the TDSC-NP-WiMAX and LTE methods and the reference value for $P_{FA}$=0.01, CP length = 1/4

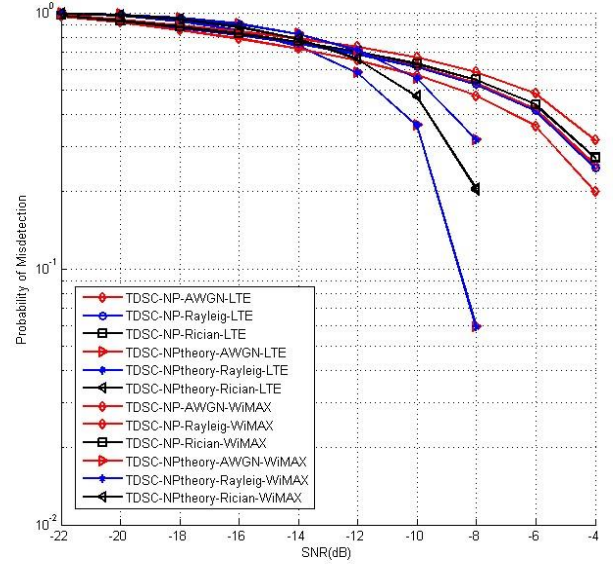

Figure 11. Performance Comparison of the TDSC-NP-WiMAX and LTE methods and the reference value for $P_{FA}$=0.01, CP length = 1/8

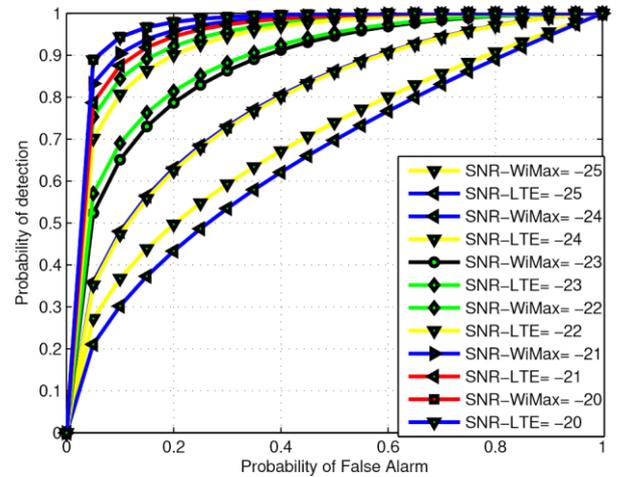

Figure 12. WiMax and LTE ROC curves for TDSC method for cyclic prefix equal to 1/4

Furthermore, it is evident that the SNR performance exhibits similar curves for both simulation and theoretical results. This similarity can be attributed to the inherent characteristics of the AWGN channel.

In Figure (11), the simulation results of Receiver Operating Characteristic (ROC) curves for the TDSC method are presented for an OFDM system with a cyclic prefix length of 1/4 using both WiMAX and LTE techniques. The graph illustrates that the OFDMA system employed by LTE shows identical performance to the WiMAX technique.

## X. CONCLUSIONS

In conclusion, this study examined the performance of the Time-Domain Symbol Cross-Correlation Non-Parametric (TDSC-NP) method for spectrum sensing in WiMAX systems. The study also suggests future investigations to explore the performance of the TDSC method in both WiMAX and LTE systems, as well as a comparison of spectrum sensing between WiMAX and LTE.

The results demonstrated that the TDSC-NP method exhibited favorable performance at low signal-to-noise ratios (SNRs) for both WiMAX and LTE systems. The simulations were conducted for different cyclic prefix (CP) ratios, including 1/4 and 1/8. The performance evaluation encompassed various channel conditions, including AWGN, Rayleigh, and Ricean fading.

Considering the similarities between WiMAX and LTE, such as their utilization of Orthogonal Frequency Division Multiple Access (OFDMA) in the downlink, higher order modulation and coding schemes, support for both Frequency Division Duplexing (FDD) and Time Division Duplexing (TDD), and compatibility with higher order MIMO antenna solutions, it is observed that both systems exhibited comparable sensing behavior. This finding is supported by the simulation results presented in the figures.

In addition, the TDSC-NP method showed promising performance in spectrum sensing for WiMAX systems, and further research is warranted to explore its application in LTE systems. The study also highlights the similarities between WiMAX and LTE in terms of their capabilities and potential for efficient spectrum utilization. Overall, this study provides valuable insights into the performance characteristics of WiMAX and LTE in different channel conditions, highlighting their strengths and potential applications. The findings can assist in the design and optimization of wireless communication systems and contribute to the advancement of future wireless technologies.